# TAMP-OS: An Open-Source Workflow for Tactile 3D-Printable Lithographs

Natalia Gonzalez-Vazquez[1+], Robert Faulkner[2+], Victoria Gamez[3,4], Karly E. Cohen[5], Gunther Richter[1], Abigale Stangl[3,4*], Andrew K. Schulz[4,6*]

[1]Materials Centralized Scientific Facility, Max Planck Institute for Intelligent Systems (MPI-IS), Stuttgart, Germany; [2]Robotics Centralized Scientific Facility, MPI-IS, Stuttgart, Germany; [3]School of Industrial Design, Georgia Institute of Technology, Atlanta, GA; [4]Tactile Media Alliance, Atlanta, GA; [5]Friday Harbor Laboratories, University of Washington, Friday Harbor, WA; [6]Haptic Intelligence Department, MPI-IS, Stuttgart, Germany;

*+These authors contributed equally.*

**Co-corresponding authors: Abigale Stangl (abigale.stangl@design.gatech.edu); Andrew K. Schulz (aschulz@is.mpg.de)*

## ABSTRACT

Describe an animal without using the verb look. The world is filled with features too small for our eyes to see: the setae on a gecko's feet, the cuticles covering a rat's whisker, or the fuzziness of a bat's wing. Can you effectively provide an alternative method for interpreting complex microscopy images while preserving the length scale? In this work, we provide a fully open-source lithograph workflow, allowing rapid creation of 3D printable lithograph files for tactile accessibility for generalized images. The lithographs made in this workflow utilize a customizable API, allowing users to manually input various output variables (resolution, relief height, and blur) while outputting various file types (STL, STP, 3MF, and GLB). This workflow is validated on several different commercial 3D printers. This work seeks to leverage 3D printing to create tactile graphics and art, making science more accessible and enabling novel tactile exploration of biological structures. This workflow is paired with an open-source GitHub repository for Tactically Accessible Microscopy Printing (https://github.com/nagova/TAMP-OS), or TAMP-OS. The TAMP-OS workflow includes a step-by-step guide, allowing other users to branch and augment the code for their customizable use cases while maintaining an open-source codebase allowing users to adapt all parts of the workflow for their needs.

## INTRODUCTION

Very few fields have advanced as rapidly as microscopy (Denk et al. 1990; Huisken et al. 2004; Pawley and Masters 2008; Ghaziyani et al. 2018; Arai et al. 2022; Bond et al. 2022; Botifoll et al. 2022; Chua et al. 2022; Aghigh et al. 2023; Quarta et al. 2023). Microscopy continues to advance, enabling imaging of single molecules(Denk et al. 1990; Mohandas and Gallagher 2008; Anggara et al. 2023), and each year, improvements allow imaging of ever smaller features. Microscopy has enabled physicists, engineers, biologists, neuroscientists, and more to capture images ranging from a single red blood cell at 2 micrometers to complex biological tissue slices at the microscale, and even 3D volumetric images using scanning electron microscopy (Denk and Horstmann 2004; Bushby et al. 2011). These techniques have become so advanced that Nikon now hosts microscopy image and photography competitions in its Photomicrography collection(Bushby et al. 2011; Li 2022), as do scientific societies (Lyman 2023). Many of these journals and magazines feature stunning cover images that showcase the beauty of the biological world, captured either through a camera lens or a microscope.

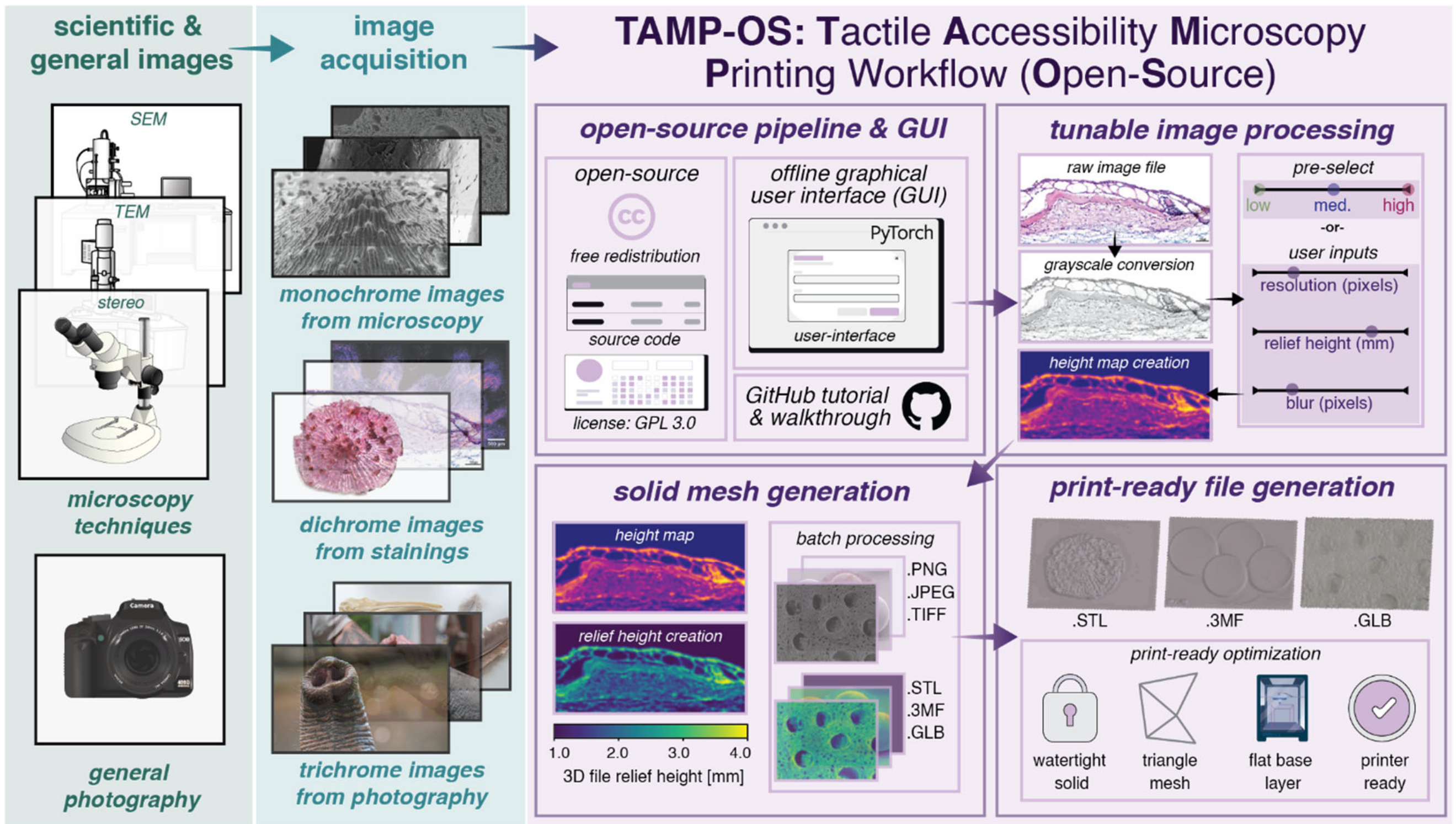

**Figure 1. Workflow for TAMP-OS: an open-source workflow for tactile 3D printable lithographs**. This workflow showcases the use of a range of microscopy techniques, including SEM, TEM, light microscopy, and confocal microscopy, and the conversion of their images into 3D files through image processing. This is the open-source workflow discussed in the Tactile Accessibility Microscopy Printing (TAMP-OS) GitHub repository associated with this paper (https://github.com/nagova/TAMP-OS).

With such remarkable imagery comes an opportunity to explore how images, graphics, and art can be made accessible through another sense—touch. Conferences on human-computer interaction, such as Computer Human Interaction, require alternative text ("alt-text") for images(Hersh and Johnson 2008; "Accessibility Guide for Authors" 2026), which explains the content and meaning of a visual, graphic, or image for people who cannot see it, making it a core part of accessible design. Scientific documents often rely on graphics alongside written text to communicate information, but these materials are frequently difficult for blind and low-vision readers to access. A large share of scientific PDFs still lacks usable author-generated alt text (e.g.,(Hersh and Johnson 2008; Wang et al. 2021; Chintalapati et al. 2022; Sirokman et al. 2023)). All the while, alt-text cannot always capture the spectral patterns of a feather's rachis-barb-barbule structure, whereas touch alone allows researchers to distinguish a songbird feather from an owl's, owing to their distinct mechanical properties (Bonser 1996; Bachmann et al. 2012). Just the same, alt text cannot convey the structural gradients found in microCT 3D scans of biological material, where enlarged tactile representations support greater understanding (Götzelmann 2018; Sirokman et al. 2023). The natural world–visible or invisible to the naked eye–is rich with tactile texture (Bonser 1996; Bachmann et al. 2012).

Tactility in design and art can be created in many ways ("ITD Journal: Tactile Graphics: An Overview And Resource Guide" 2026; Edman n.d.), including through traditional arts and craft practices and materials such as collage, hand modeling, knitting, and raised-line forms; through specialized materials or production systems such as embossing, swell paper, or microcapsule processes, and embroidery (e.g., (Ichihashi et al. 2025; Seehorn et al. 2025)). It can also be achieved through digital approaches such as 3D modeling and additive manufacturing, as well as emerging automated computational design and manufacturing techniques (e.g.,(Panotopoulou et al. 2020; Clark and Pavel 2025; Marriott et al. 2025; Kukreja and Jain 2026)). Manual and technical processes, including embossing technologies, swell-paper production, and textile-based methods, and additive manufacturing or three-dimensional printing (3D printing), offer yet another avenue of tactile reproductions (e.g., (Ichihashi et al. 2025; Seehorn et al. 2025)). Each method has particular strengths, such as supporting multisensory learning or making spatial relationships more interpretable. Yet, their adoption in some academic fields as modes of communication has been slow (Rong et al. 2025). This is due to the need for specialized machinery, complex production processes, and limited awareness of available methods and resources (Marriott et al. 2025; Rong et al. 2025).

Notably, 3D printing has emerged as an increasingly affordable pathway to tactile accessibility (Sirokman et al. 2023), most commonly from *STL* or *CAD* files, captured volumetric data such as microCT reconstructions, and, more broadly, tactile graphics and accessible physical representations. Literature on workflows for 3D-printing tactile graphics generally describes a multi-stage process rather than a single standardized method, typically involving source selection, tactile abstraction of visual content, digital modeling, fabrication, and evaluation with blind or low-vision users for usability and comprehension (Wu et al. 2022; Rong et al. 2025). Across this work, the critical step is not printing alone but the deliberate conversion of visual structure into tactile form, with design decisions around representation type, scale, and complexity shaping whether the final artifact is legible and useful(Wu et al. 2022). Accessible authoring and participatory refinement are also increasingly central to this workflow, as projects such as *shapeCAD* shift the process from designing only for blind users toward designing with and by them (Siu et al. 2019), while newer systems demonstrate a production workflow that automatically processes an image for automated fabrication of multiple replicas of that content in different material or form (Clark and Pavel 2025). More broadly, tactile-graphics design is an active area of translation between visual and tactile representations, while also highlighting ongoing challenges related to standards, design knowledge, and adoption (Marriott et al. 2025; Rong et al. 2025).

Building on this literature, this work extends tactile-graphics workflows into scientific imaging by presenting an open-source pipeline that takes printer specifications as co-design features for printing both microscopy images (e.g., SEM, TEM, SHG, light microscopy, and staining images) or general photography/images into 3D files of various types (e.g., STL, STP, 3MF, GLB) which can be 3D-printed. Our workflow includes a step-by-step guide through a fully open-source GNU 3.0 licensed GitHub ([https://github.com/nagova/TAMP-OS](https://github.com/nagova/TAMP-OS)) allowing for rapid (< 30s) batch-production of 3D files. Similar to prior work on tactile representation, this approach treats tactile translation not simply as a production step but as an accessibility-centered design process that expands the ways visual information can be interpreted and explored. The workflow is fully open-source and customizable allowing for users to customize all the source code for their specific user needs. The goal of this manuscript and workflow is to make chemistry, biology, materials science, neuroscience, and adjacent disciplines more inclusive by enabling tactile exploration not only as a mode of access but also as a possible mode of scientific discovery for disabled individuals.

## METHODS

### Creation of the TAMP workflow

The TAMP pipeline was implemented in Python using NumPy, Pillow, SciPy, and numpy-stl, with no proprietary dependencies, and distributed as a series of self-contained Jupyter notebooks for reproducible batch processing and parameter optimization. Input microscopy images were first preprocessed in ImageJ/Fiji (Schindelin et al. 2012) to remove metadata bars, scale bars, and annotations that would otherwise appear as raised topographic features in the final lithograph. Images were exported as PNG or TIFF files for downstream processing. Within the pipeline, images were converted to grayscale if not already preprocessed, followed by contrast stretching (2nd–98th percentile clipping) and optional Gaussian blur filtering to suppress high-frequency SEM noise or fine structural detail below the printable resolution threshold. Pixel intensities were then normalized to a [0, 1] floating-point range and vertically inverted to preserve tactile orientation before height map generation.

Watertight mesh construction combined the triangulated top-relief surface with a flat base and four enclosing side walls. Aspect ratio was automatically preserved during print scaling, with a warning system flagging any user-defined dimensions that would introduce geometric distortion. Users could define print dimensions manually or allow the software to calculate them from the original image geometry. Output meshes were exported in STL, 3MF, and GLB formats. Hardware-aware parameter presets were derived from nozzle diameter and layer height to improve reproducibility across diverse 3D printing technologies. The Low, Medium, and High presets scale lateral image resolution to custom mesh outputs, having finer meshes for higher resolutions. All parameters, including resolution, blur radius, base thickness, and relief height, are fully overridable to accommodate different microscopy modalities and fabrication platforms.

A parameter-sweep notebook was developed to systematically evaluate the effects of lateral resolution, Gaussian blur, and relief height on tactile surface morphology by varying one parameter at a time across a single representative image. Generated STL files were visualized using automated PyVista rendering, producing standardized full-view and close-up screenshots for quality assessment and figure preparation.

### TAMP visualization and parameter workflow

To support systematic optimization of lithograph parameters prior to committing to a full batch run, the TAMP toolbox includes two companion tools: a parameter sweep notebook and an automated rendering notebook. The parameter sweeps notebook processes a single representative image through the full pipeline across a user-defined range of one parameter at a time — lateral resolution, relief height, or Gaussian blur — while holding all others fixed. For each value tested, a clearly named output file is generated and a side-by-side height map preview is saved as a PNG, allowing direct visual comparison of how each parameter affects surface morphology before any physical printing is performed. Tested ranges include lateral resolution (64–512 pixels), relief height (1.00–5.00 mm), and blur sigma (0.5–2.0 pixels), with all other pipeline parameters remaining at user-defined defaults throughout the sweep.

The automated rendering notebook addresses the need to visually inspect batches of generated STL files without manually loading each one into a mesh viewer. Given a folder of STL files, the notebook renders each mesh automatically in two views — a full isometric view and a zoomed close-up — using PyVista with orthographic projection and a two-light setup to produce clean, matte-shaded renders suitable for manuscript figures and quality assessment. Camera angle, zoom, and background theme are fully configurable, and all renders are saved as PNG files to separate output folders without requiring any manual interaction.

## Open-sourcing statement for TAMP

The TAMP toolbox is distributed as open-source software under the GNU General Public License version 3.0 (GPL-3.0), an OSI-approved license that satisfies all ten criteria of the Open-Source Definition ("The Open Source Definition" 2006) maintained by the Open-Source Initiative (OSI). The source code is made available in full and in its preferred form for modification; redistribution is unrestricted and free of royalty requirements; and derived works and modifications are explicitly permitted, provided they are distributed under the same GPL-3.0 terms — a copyleft condition that ensures downstream versions of the toolbox remain openly accessible to the research community. No restrictions are placed on any person, group, or field of endeavor, including commercial or research use, and the license is technology-neutral, attaching rights to the software itself rather than to any particular distribution or platform. All dependencies — NumPy, Pillow, SciPy, numpy-stl, and optionally PyVista — are themselves OSI-compliant open-source packages, satisfying the OSD requirement that the license not place restrictions on co-distributed software. The toolbox requires no proprietary software at any stage of the pipeline: image preprocessing, mesh generation, slicing, and printer control can all be performed with freely available, community-maintained tools. This design ensures that the full workflow is reproducible, auditable, and extensible by any researcher without cost barriers or hardware lock-in, consistent with the principles underlying the OSD.

## Microscopy images utilized for TAMP pipeline and Lithograph printing

Microscopy images were drawn from open-source, previously published datasets, including images of lumpsucker armor(Woodruff et al. 2022; Hoover et al. 2023), gold nanowhiskers(Huang et al. 2019), Janus particles,(Gonzalez-Vazquez et al. 2026) elephant whiskers(Schulz et al. 2026), rodent teeth(Srot et al. 2024), and elephant skin collagen(Schulz et al. 2025). All raw images were sourced from publicly available data files, with citations provided at each point of reference in the manuscript. The images shown in Fig. 1 – Fig 4 were selected to highlight differences in texture, layering, morphology, and structure across biological samples at varying length scales, demonstrating the breadth of the workflow's applicability.

### Printing specifications for Lithographs

Lithographs were produced on multiple printers to assess the quality and reproducibility across a variety of technologies for TAMP-OS including both prints by the authors and beta-testing persons for the TAMP-OS repository. Stereolithography (SLA), multijet printing (MJP), and fused deposition modeling (FDM) (Karakurt and Lin 2020) were all used to test and validate the workflow. Printers included the Stratasys J835 (MJP), Carbon M2 (SLA), and Bambu Labs X1E (FDM). Materials included liquid photopolymer resins for the J835 and Carbon M2 (Vero Black and Loctite 3843), while polylactic acid, better known as PLA, was used for the Bambu (FDM). Print settings included 0.12mm layer height with a 0.4mm nozzle for the Bambu FDM print, the J835 used a 0.027mm layer thickness, 82 grams of material; lastly, the Carbon M2 used 0.1mm layers, 45 mL of material.

## RESULTS

The open-source workflow successfully generated tactile lithographs from microscopy images spanning multiple imaging modalities, including SEM, TEM, bright field microscopy. Across all tested image types, the pipeline reproducibly converted microscopy contrast gradients into discrete extruded surface topographies suitable for FDM fabrication (**Fig. 2**). A parameter sweep demonstrated that tactile feature quality depended strongly on the relationship between image resolution, Gaussian blur, and printer nozzle diameter, with excessive resolution producing unnecessarily large meshes without improving physically printable details. Medium-resolution presets, corresponding approximately to one pixel per nozzle width, consistently provided the best balance between tactile fidelity, file size, and print reproducibility across tested systems. The workflow additionally enabled rapid comparative visualization of lithograph quality through automated rendering of generated meshes. Systematic sweeps of blur radius, relief height, and lateral resolution demonstrated predictable effects on tactile surface morphology. Increased blur reduced high-frequency surface noise in SEM-derived lithographs, while increased relief height enhanced tactile contrast at the expense of steeper surface gradients. Resolution sweeps showed diminishing returns once the encoded feature size exceeded the physical resolving capability of the printer nozzle.

Importantly, the workflow remains compatible with low-cost FDM systems. The generated geometries were successfully exported in multiple standard mesh formats and fabricated across printers utilizing FDM, SLA, and multijet technologies. This demonstrates that the workflow is not tied to a single hardware ecosystem and can serve as a generalized framework for accessible tactile fabrication of microscopy imagery.

### TAMP Batch GUI: A Hardware-Aware Interface for 3D Lithograph Generation

The TAMP toolbox converts microscopy images into 3D-printable lithographs through a four-stage pipeline: image preprocessing, height map generation, mesh construction, and file export (see **Fig. 2**). Input images (PNG, JPEG, or TIFF) are converted to grayscale and contrast-stretched using percentile clipping (2nd–98th percentile) to normalize intensity across varying acquisition conditions. A Gaussian blur suppresses high-frequency noise that would otherwise produce unprintable surface artifacts, and pixel values are normalized to a [0, 1] float range before mesh construction. The resulting mesh is a fully watertight solid combining a triangulated top-relief surface, a flat base layer, and four closing side walls, ensuring direct compatibility with standard FDM slicers without manual mesh repair. Output files are exported in STL, 3MF, and GLB formats as selected by the user (**Fig. 2**).

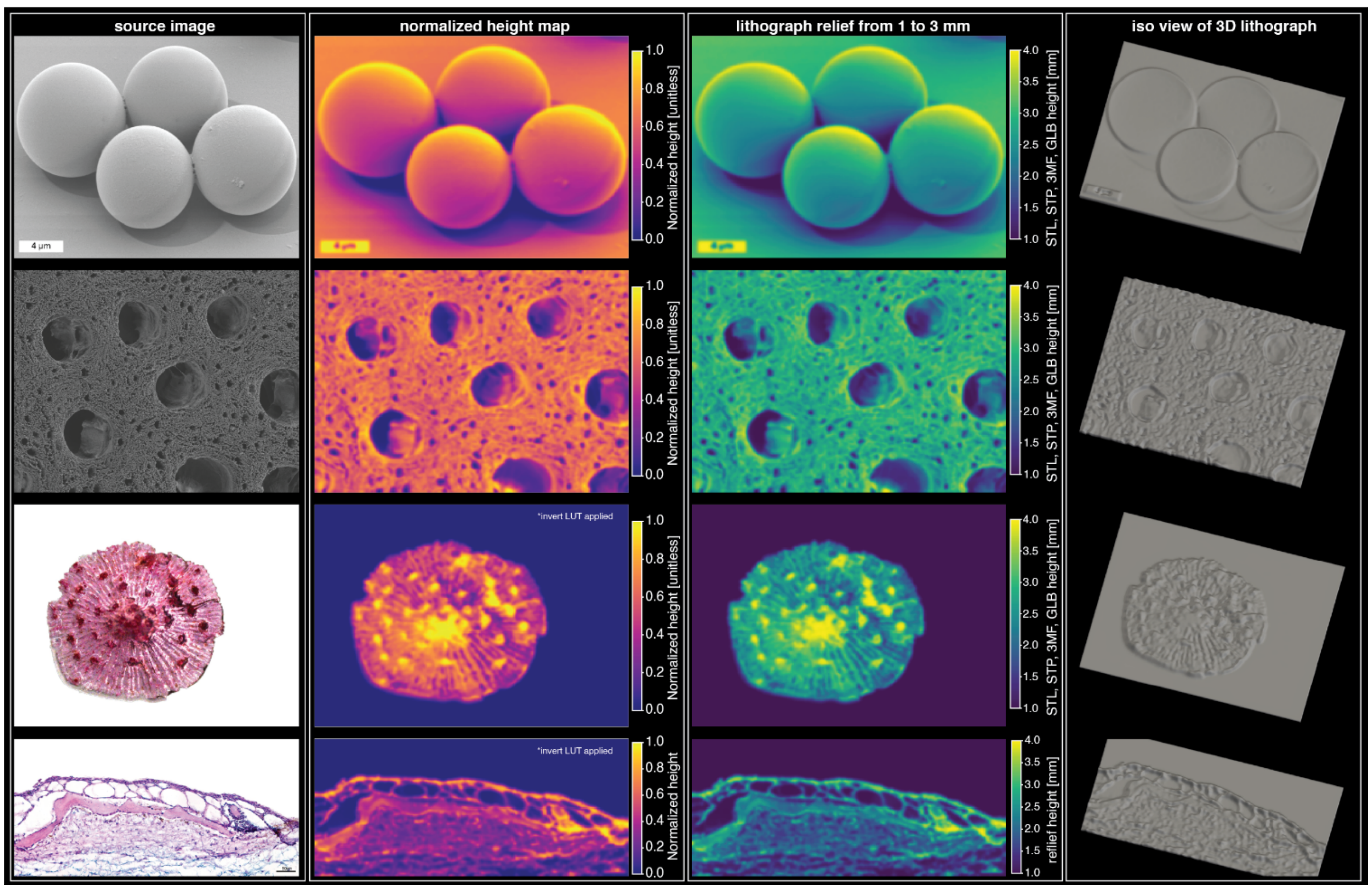

**Figure 2. Example outputs from the TAMP-OS workflow**. This workflow showcases the use of a range of microscopy techniques, including SEM and light microscopy. For each image we show the source image, a normalized height map transformation of the image, the creation of the lithograph relief, and an isometric view of the 3D lithograph that is available in STL, STP, 3MF, and GLB exportable formats. For the bottom two rows of light microscopy images the colors are inverted prior to height map creation.

The primary user interface is the TAMP Batch GUI (v2), a graphical application launched from a Jupyter notebook that enables batch conversion of large image folders in a single session (**Fig. 1**). A central feature is the hardware-aware preset system, which derives resolution and blur parameters directly from the user's printer specifications — nozzle diameter and layer height — rather than requiring manual tuning. Three presets are available: Low (one pixel = twice the nozzle diameter), Medium (one pixel = one nozzle diameter, recommended default), and High (one pixel = half the nozzle diameter, capturing sub-nozzle texture at the cost of larger file size). For a standard 0.4 mm nozzle on a 100 mm wide print, these correspond to resolutions of 128, 256, and 512 pixels, respectively. The interface also displays the number of discrete height levels available at the user's specified relief and layer height, providing a concrete estimate of vertical resolution prior to file generation. A full customization panel allows independent override of resolution, blur, and base thickness for users requiring finer control.

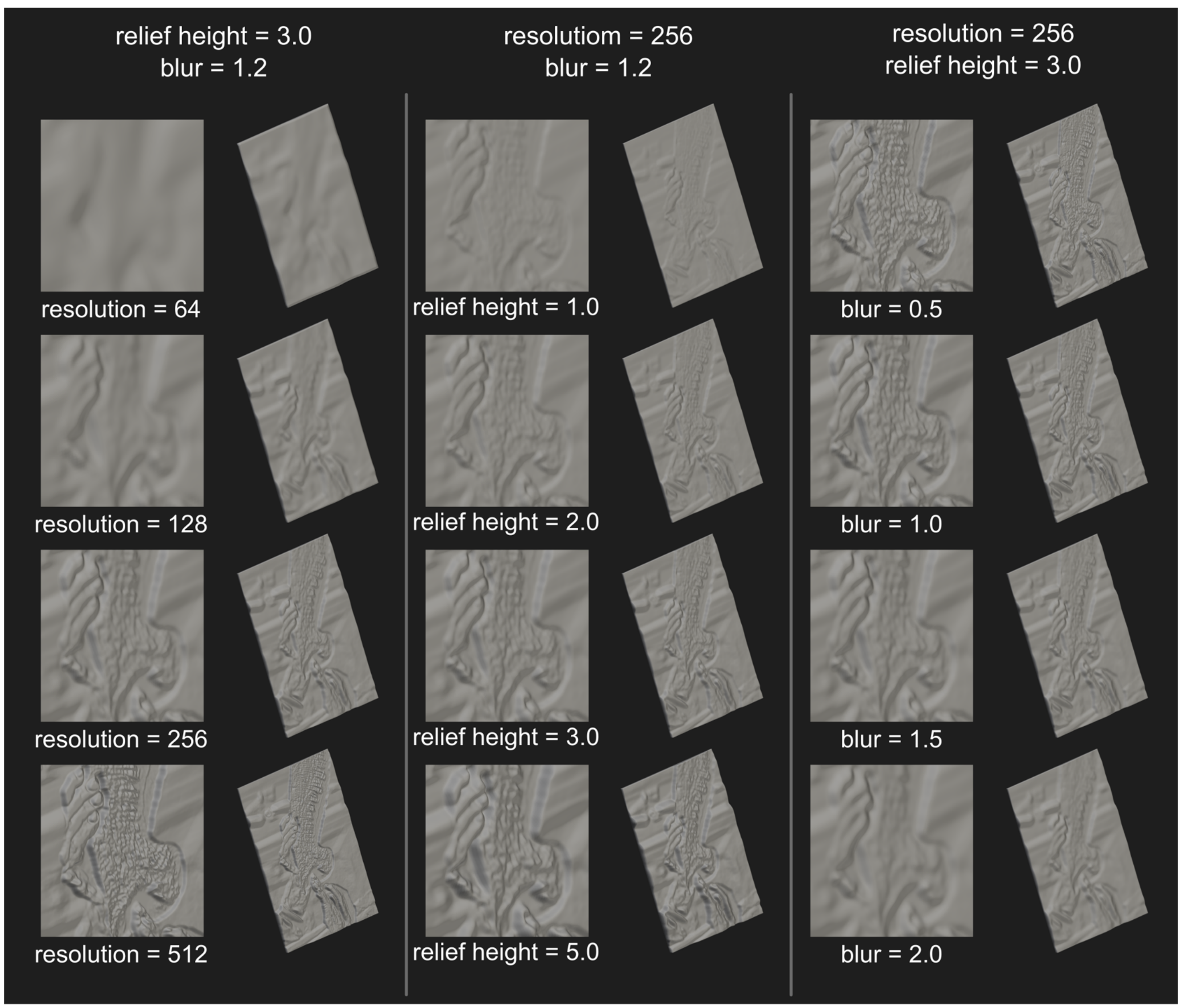

**Figure 3. Customizability of the TAMP-OS workflow**. This workflow showcases the use of a range of microscopy techniques, including SEM, TEM, light microscopy, and confocal microscopy, and the conversion of their images into 3D files through image processing

To support parameter optimization prior to a full batch run, the toolbox includes a companion resolution comparison notebook that sweeps one variable — resolution, relief height, or blur — across a single representative image while holding all others fixed. A separate visualization notebook uses PyVista to render all STL files in a target folder as isometric and close-up screenshots without manual interaction, enabling direct visual comparison of parameter effects. Together, these components provide a complete and reproducible workflow from raw microscopy image to print-ready file, compatible with any FDM printer running open-source firmware (e.g., Klipper, Marlin) and any open-source slicer (e.g., PrusaSlicer, OrcaSlicer, Cura).

### Stress testing TAMP workflow

As the TAMP workflow works for various files, we worked on stress testing the workflow by varying three of the variables that can be pre-set or customized by the user: relief height, blur, and resolution. This stress testing was first done for the lithograph creation to look at how the different features affect the output images as is shown by Fig. 3.

As shown by Fig. 3 the relief height changes the relief of the texture making it more visually apparent on the lithograph 3D renderings whereas the blur changes appear to introduce artifacts around the more non-linear portions of the color-map. The resolution for the images stress tested dramatically reduces the file size, yet simultaneously reduces nearly all of the features on the lithograph.

### Validation of the TAMP workflow on multiple printer types

The workflow supports both single images and multi-panel figures. To evaluate the workflow, we used microscopy images from previously published, open-source datasets, including images of lumpsucker armor, elephant whiskers, elephant skin collagen, and others (details included in the methodology). The workflow was tested across a range of imaging modalities and complexities, including scanning electron microscopy (SEM), transmission electron microscopy (TEM), second-harmonic generation (SHG) microscopy, and brightfield light microscopy. The workflow reliably produced a diverse set of lithographs with a SEM lithograph demonstrated in **Fig. 4.** The 3D files were evaluated across three printer technologies: SLA, MJP, and FDM.

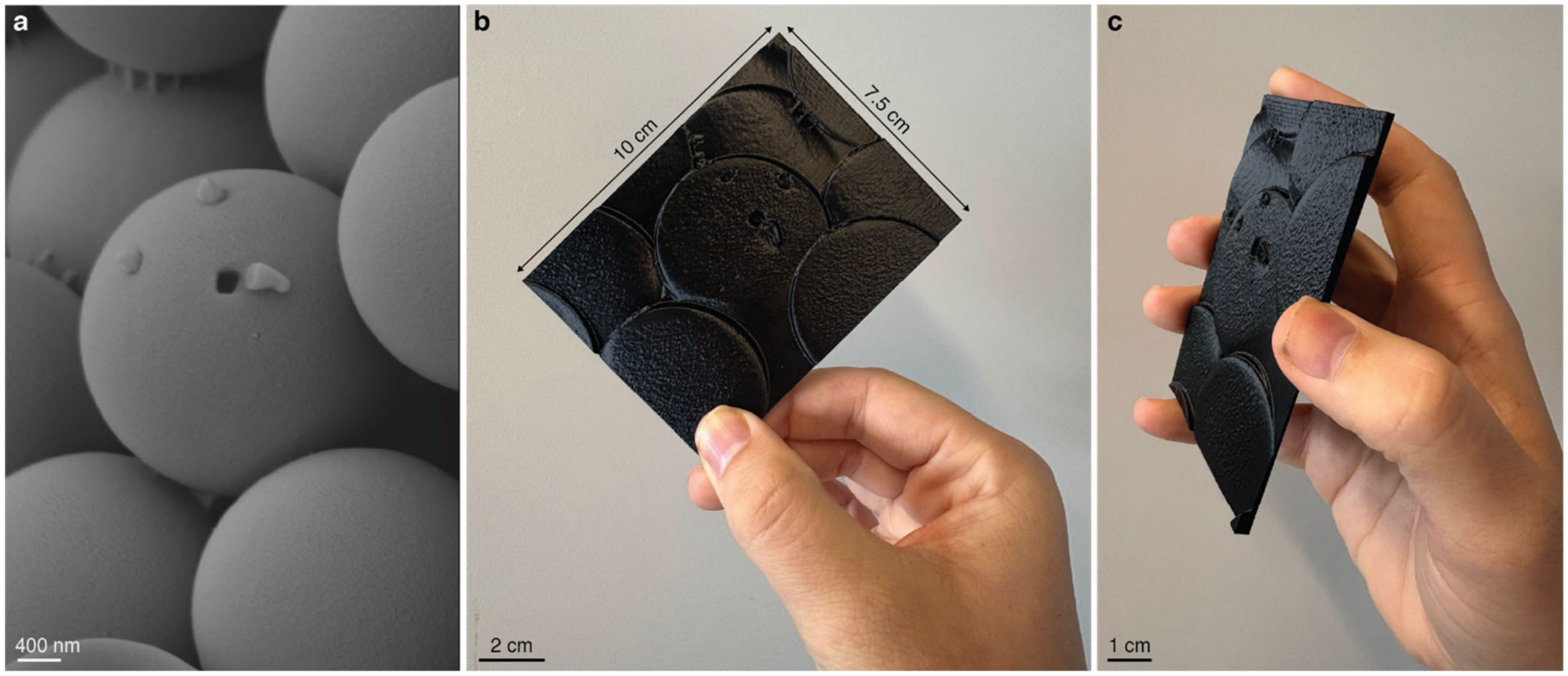

**Figure 4.** Demonstration of the TAMP workflow with different views from Vazquez et al.(Gonzalez-Vazquez et al. 2026). a) SEM image of a nanometer scale particle. b) Top-down view of the 7.5 cm x 10 cm 3D printed lithograph of the SEM image in a. C) Isometric view of the 3D printed lithograph of the SEM image in a.

To evaluate the workflow, we used microscopy images from previously published, open-source datasets, including images of lumpsucker armor, elephant whiskers, elephant skin collagen, and others (details included in the methodology). The workflow was tested across a range of imaging modalities and complexities, including scanning electron microscopy (SEM), transmission electron microscopy (TEM), second-harmonic generation (SHG) microscopy, brightfield light microscopy, as well as generalized scientific and non-scientific photographs (**Fig. 1** – **Fig. 4**). As shown by Fig. 4 this workflow allows taking SEM images of items at the length scale of 100s of nanometers (Fig. 4a) and to generate tactile lithographs at the centimeter length scale (Fig. 4b-c).

## DISCUSSION

The workflow shared for 3D-printing lithographs for the digital accessibility of microscopy images is important because it offers a practical alternative to prior methods for offline production of tactile microscopy images (e.g., (Koone et al. 2022)). While 3D printing can succumb to many challenges/failure modes when printing high-detailed 3D files (Baechle-Clayton et al. 2022) the TAMP-OS workflow allows customization of different lithograph creation features for adjustments if artifacts should be introduced, or relief height is not enough for the specific use-case. Many previous approaches for lithograph creation are not purely open-source with customizable APIs as they are browser-based pipelines, or image-to-height-field approaches that break down under the density and contrast complexity of microscopy data. Earlier lithography and lithophane workflows helped demonstrate that image-based tactile rendering could broaden access to visual material (e.g.,(Romat and Jamaludin 2022; Avni et al. 2026)), but they have generally worked best for simpler images or have required levels of print precision that place them out of reach for most laboratories, classrooms, and public-facing scientific institutions.

In contrast, the workflow presented in this paper was developed specifically for the realities of microscopy: high-resolution imagery, fine structural variation, dense contrast, and the need for a process that can be implemented without specialized fabrication infrastructure. By an open-source workflow with the cost accessibility of 3D printing, this approach makes tactile microscopy more achievable in research and educational settings allowing a 10 by 10 cm print to be made for under one dollar (**Fig. 4b-c**). The practical significance of this contribution lies in a workflow that is modular, open-source, and low-cost, which makes it easier to adopt, inspect, modify, and share. Previous approaches often positioned tactile conversion as a specialized technical endpoint, dependent on either expensive machines or closed, highly constrained systems (Sarkar and Das 2012; Stangl et al. 2019). Here, scientific users can move microscopy images through a repeatable process, from source selection, image preparation, and tactile translation to fabrication, without requiring custom software development or ultra-high-end printing resources. This is a meaningful shift from prior methods because it lowers both technical and financial barriers while retaining enough flexibility to support multiple microscopy modalities, including SEM, TEM, SHG, light microscopy, and stained images.

For adoption of the workflow, the key step is to treat tactile accessibility as part of microscopy communication rather than as a downstream accommodation. In practice, this means identifying which images are central to a figure, dataset, lesson, or presentation; evaluating whether they can be meaningfully translated into tactile form; and incorporating that translation early enough to allow refinement. Scientists adopting this workflow will also need to attend to the same design questions emphasized in tactile-graphics literature more broadly (Rong et al. 2025): which features should be preserved, which should be simplified, how scale affects legibility, and when dense visual detail must be reinterpreted rather than directly mapped. The advantage of the present workflow is that these decisions can now be explored with low-cost, widely available tools rather than being limited to highly specialized environments. Below, we discuss the non-visual accessibility of this workflow as used by Blind and low-vision (BLV) scientists. This workflow is also significant in relation to Universal Design for Learning (UDL) (King-Sears 2009; Ralabate 2011), as it expands microscopy beyond a purely visual mode and supports multiple means of representation, engagement, and access. Translating microscopy images into tactile lithographs benefits blind and low-vision users while also offering educators, outreach practitioners, and scientists a more multimodal way to communicate and explore microscopic structures in classrooms, labs, museums, and conference settings. In this sense, tactile microscopy is not simply an after-the-fact accommodation, but a design strategy that broadens participation in scientific interpretation. Compared with previous methods, the contribution here is not merely that microscopy images can be printed, but that they can be translated through a workflow that is affordable, open, adaptable, and better aligned with UDL-informed accessibility goals and inclusive scientific practices.

## FUTURE WORK & CONCLUSION

This paper presents an open-source workflow and framework for converting microscopy images into digitally accessible, 3D-printable lithographs using a hardware-aware codebase. The work is ongoing, and the associated GitHub repository will be continuously updated and linked to the Tactile Media Alliance and See3D (The Ohio State University and Karbowski 2020). Future work will include user evaluation of the 3D printed tactile microscopy lithographs with blind and low-vision participants to assess tactile legibility, usability, and interpretive value across 3D printed scientific image types; comparative testing across modalities such as general photography, SEM, TEM, SHG, light microscopy. The open-source workflow includes customization of different features, such as rendering quality, blur filters, and relief height. Future work will look into the development of design guidelines for scale, contrast, simplification, and surface detail in tactile microscopy lithographs validated by user-studies. Beyond the accessibility of the end-product, future work will look at the accessibility of the workflow, in alignment with current understanding of accessible 3D Printing (Reinders et al. 2023; Ballarin et al. 2025; Zhang et al. 2025).

The authors also acknowledge that many affordable 3D printing materials carry environmental costs. Future work will explore more sustainable alternatives, including biodegradable (Lyu et al. 2021) or recycled filaments (Lodha et al. 2023), to ensure that expanding tactile accessibility does not come at the expense of environmental responsibility. Accordingly, future work will also explore more sustainable fabrication practices, including biodegradable or lower-impact filaments, while expanding the open-source repository with documentation, templates, example files, troubleshooting guides, and discipline-specific workflows to support broader adoption (Pearce 2012). Researchers across disciplines should prioritize making their work accessible beyond the visual modality. Tactile scientific exploration could yield new insights—particularly in biologically inspired design (Donatelli, Cohen, et al. 2025), where the complexity of natural materials and systems has long inspired novel technologies (Donatelli, Vandenberg, et al. 2025). Importantly, accessibility is not a one-time achievement; it must evolve alongside new tools and standards, such as advances in alt-text and neurodivergent-friendly fonts and layouts (Baxter 2025).

Future work should therefore also integrate tactile outputs with alt text and other multimodal accessibility practices, test the workflow in classrooms, museums, conferences, and public science engagement settings, improve the accessibility of the workflow itself for disabled makers and researchers, and examine how tactile microscopy might function not only as a tool for access, but also as a mode of scientific interpretation. Cross-disciplinary adoption studies and comparison with prior lithograph and tactile-graphics workflows will also be important for benchmarking cost, print quality, accessibility, and ease of adoption. The authors have intentionally kept this paper concise to encourage scientists to treat accessible, open-source scientific graphics as a standard part of their practice—tools that can be shared at conferences, networking events, and public outreach to invite broader audiences to explore the microscopic world through touch.

**ACKNOWLEDGEMENTS**

The authors are grateful for the support of the Tactile Media Alliance. A.K.S. acknowledges support from the Alexander von Humboldt Foundation. G.R. and A.K.S. acknowledge support from the Max Planck Society.

**Disclosure of Interests**

The authors have no competing interests to declare that are relevant to the content of this article.

**Data availability**

All data used in this manuscript were used with permission from the original authors. The workflow described in this paper is supported by an open-source, as defined by the open-source definition (https://opensource.org/osd), GitHub (https://github.com/nagova/TAMP-OS) repository containing all necessary dependencies and step-by-step instructions for printing lithographs from a variety of microscopy images. The images, 3D lithograph files, and examples used in the workflow and this manuscript can be found on the online Edmond repository (Gonzalez-Vazquez et al. 2026).